\documentclass{article}
\usepackage{spconf,amsmath,graphicx}

\usepackage{amsmath,amssymb,amsfonts}
\usepackage[colorlinks=true, citecolor=black, linkcolor=black]{hyperref}
\usepackage{scalefnt}
\usepackage{soul}
\usepackage{cite}
\usepackage{caption}
\usepackage{bm}
\usepackage{textcomp}
\usepackage{calc}
\usepackage[hang,flushmargin]{footmisc}
\usepackage{enumitem}
\usepackage{microtype}
\usepackage{lipsum}   
\usepackage{multicol} 
\usepackage{tabularx} 
\usepackage{setspace}
\newcommand{\ignore}[1]{}
\newcommand{\mathcalbf}[1]{\boldsymbol{\mathcal{#1}}}

\newcommand{\outprod}{\circ}
\newcommand{\rank}{\mathrm{rank}}
\newcommand{\norm}[1]{\left\| {#1} \right\|}
\newcommand{\argmin}{\mathrm{arg\,min}}

\newcommand{\tvec}{\mathrm{vec}}

\newcommand{\MATLAB}{M{\scalefont{0.85}ATLAB}}

\newcommand{\Abf}{\mathbf{A}}
\newcommand{\Bbf}{\mathbf{B}}
\newcommand{\Cbf}{\mathbf{C}}

\newcommand{\Ebf}{\mathbf{E}}
\newcommand{\Fbf}{\mathbf{F}}
\newcommand{\Gbf}{\mathbf{G}}

\newcommand{\Jbf}{\mathbf{J}}

\newcommand{\Mbf}{\mathbf{M}}
\newcommand{\Nbf}{\mathbf{N}}

\newcommand{\abf}{\mathbf{a}}
\newcommand{\bbf}{\mathbf{b}}
\newcommand{\cbf}{\mathbf{c}}

\newcommand{\fbf}{\mathbf{f}}

\newcommand{\pbf}{\mathbf{p}}

\newcommand{\rbf}{\mathbf{r}}

\newcommand{\zbf}{\mathbf{z}}

\newcommand{\Ccalbf}{\mathcalbf{C}}

\newcommand{\Ncalbf}{\mathcalbf{N}}

\newcommand{\Pcalbf}{\mathcalbf{P}}
\newcommand{\Qcalbf}{\mathcalbf{Q}}

\newcommand{\Tcalbf}{\mathcalbf{T}}

\newcommand{\Wcalbf}{\mathcalbf{W}}
\newcommand{\Xcalbf}{\mathcalbf{X}}
\newcommand{\Ycalbf}{\mathcalbf{Y}}
\newcommand{\Zcalbf}{\mathcalbf{Z}}

\newcommand{\Hcal}{\mathcal{H}}

\newcommand{\Lcal}{\mathcal{L}}

\newcommand{\Pcal}{\mathcal{P}}

\newcommand{\CC}{\mathbb{C}}

\newcommand{\LL}{\mathbb{L}}

\newcommand{\VV}{\mathbb{V}}

\ignore{
\tolerance=4000
\emergencystretch=9pt
}

\title{A Block Term Decomposition Model Based Algorithm for Tensor Completion of Multidimensional Harmonic Signals}
\name{Lei Wang, Xiao-Feng Gong, Xi-Yuan Liu, Wei Feng, and Qiu-Hua Lin%
\thanks{This work is supported by National Natural Science Foundation of China under grants 62471084, 62471082, 62071082, and 61871067, and China Postdoctoral Science Foundation under grant 2020M680922.}}%
\address{School of Information and Communication Engineering, Dalian University of Technology, China\\E-mails: xfgong@dlut.edu.cn}

\begin{document}
\maketitle
\begin{abstract}
We consider tensor data completion of an incomplete observation of multidimensional harmonic (MH) signals. Unlike existing tensor-based techniques for MH retrieval (MHR), which mostly adopt the canonical polyadic decomposition (CPD) to model the simple “one-to-one” correspondence among harmonics across difference modes, we herein use the more flexible block term decomposition (BTD) model that can be used to describe the complex mutual correspondences among several groups of harmonics across different modes. An optimization principle that aims to fit the BTD model in the least squares sense, subject to rank minimization of hankelized MH components, is set up for the tensor completion task, and an algorithm based on alternating direction method of multipliers is proposed, of which the effectiveness and applicability are validated through both numerical simulations and an application in Sub-6GHz channel state information (CSI) completion.
\end{abstract}
\section{Introduction}
\label{sec:intro}

\ignore{
Multidimensional harmonic retrieval (MHR) is a fundamental problem in signal processing \cite{Schur_MHR,CA_VAI,CPD_array,Tensor_MHR_MIMOradar,HOSVD_MHR,VDM_CPD,MHR_CCPD_2,RecALS,HankMatNN_TC,DOA_Coarray_MS,CoarrayDOA,CTC_CPD}. Recently, tensor-based approaches to MHR have attracted considerable attention \cite{CPD_array,Tensor_MHR_MIMOradar,HOSVD_MHR,VDM_CPD,RecALS,MHR_CCPD_2,HankMatNN_TC} and have found extensive applications in areas such as array signal processing \cite{CPD_array,HOSVD_MHR,VDM_CPD,MHR_CCPD_2,DOA_Coarray_MS,CoarrayDOA,CTC_CPD}, MIMO radar \cite{Tensor_MHR_MIMOradar}, blind source separation \cite{VDM_CPD} and high-dimensional spectroscopy \cite{HankMatNN_TC}. 
Compared with matrix-based methods, tensor-based methods leverage the data more effectively while preserving the correspondence among harmonic components across different modes, resulting in improved accuracy.
}

Multidimensional harmonic retrieval (MHR) is a fundamental problem in signal processing \cite{Schur_MHR,Tensor_MHR_MIMOradar,HOSVD_MHR,MHR_CCPD_2,RecALS,ESPRIT_MIMO,HankMatNN_TC,8474369,2D_CS,CAI2016470,LowRank_Hankel}. Recently, tensor-based approaches to MHR have attracted considerable attention \cite{Tensor_MHR_MIMOradar,HOSVD_MHR,RecALS,MHR_CCPD_2,ESPRIT_MIMO,HankMatNN_TC}\linebreak and have found extensive applications in areas such as array signal processing \cite{Tensor_MHR_MIMOradar,HOSVD_MHR,MHR_CCPD_2}, channel estimation \cite{ESPRIT_MIMO} and high-dimensional spectroscopy \cite{HankMatNN_TC}. 
Compared with matrix-based methods, tensor-based methods leverage data more effectively and preserve the correspondence among harmonics across different modes, resulting in improved performance.

In existing researches, harmonic data completion from incomplete observations has gained much attention, where the incomplete signals may arise from costly experiments, equipment limitations, or sparse sampling \cite{HankMatNN_TC,8474369,2D_CS,CAI2016470,LowRank_Hankel}. Methods\linebreak in \cite{8474369,2D_CS,CAI2016470,LowRank_Hankel} make use of the low-rank property of hankelized/\linebreak toeplitzized harmonics, and propose rank-minimization based methods for data completion. Tensor data completion works in \cite{tomasi2005parafac,ACAR201141,BTDCOM_rank,TTCom,TRCom} mainly exploit the multilinear structure of the partially observed data, and have been extensively applied in image and video completion tasks. The work in \cite{HankMatNN_TC} considered MHR data completion based on canonical polyadic decomposition (CPD) model, where the CPD structure and the low rank property of harmonics are jointly exploited via a ``model fitting plus rank minimization" multi-objective optimization procedure.

\ignore{
In existing researches, multidimensional harmonic (MH) data completion from incomplete observations has gained much attention, where the incomplete signals may arise from costly experiments, equipment limitations, or sparse sampling \cite{HankMatNN_TC,com2DHR,DOA_Coarray_MS,CTC_CPD,CA_VAI}. Methods in \cite{com2DHR,CA_VAI} perform completion by leveraging the properties of the harmonic components on matrix-based models via optimization technologies; while \cite{CTC_CPD,DOA_Coarray_MS} construct tensors from the canonical polyadic decomposition (CPD) model data by utilizing the harmonic structure, and conduct completion based on its low-rank structure rather than the inherent multilinear structure. Besides, though many tensor completion algorithms \cite{tomasi2005parafac,ACAR201141,BTDCOM_rank,TTCom,TRCom} effectively leverage the multilinear structures for completion, especially in image and video tasks, they tend to ignore the harmonic structure. Instead, the work in \cite{HankMatNN_TC} considers both aspects on the CPD model.
}
\ignore{
Besides, MHR based on mutilinear rank (MLR)-$(L_r,M_r,\cdot)$ block term decomposition (BTD) model, which is found to be equivalent to the so-called double coupled CPD (DC-CPD) model \cite{GONG2022108716} is considered in \cite{Tensor_MHR_MIMOradar}.
}

Nevertheless, the above-mentioned CPD based MH data completion researches typically assume a simple ``one-to-one''\linebreak correspondence between harmonics across different modes, i.e., each harmonic in a certain mode interacts with precisely one harmonic in each other mode \cite{HankMatNN_TC}. However, in practice, harmonics in different modes often exhibit a more complex mutual correspondence. For instance, according to the standard 3GPP communication protocol \cite{3GPP}, under the Sub-6GHz spectrum, the MHR model for the wireless channel state information (CSI) in ``space-time-frequency" domain is featured with the characteristics that each harmonic in frequency domain is associated with a ``cluster", and thus interacts with a number of ``space-time" harmonics that are associated with ``paths" in that cluster, hence a ``one-to-many" correspondence among harmonics in different modes in this MHR model. As such, the CPD model is incapable of well defining such complex interactions among MH components, leading to performance loss if the associated data completion techniques are adopted. Therefore, methods based on a new tensor model need to be investigated, which can\linebreak better adapt to these complex-corresponding MHR scenarios.
\ignore{
However, existing research typically assumes a simple ``one-on-one'' correspondence among harmonic components across different modes, which is modeled using the CPD model \cite{HOSVD_MHR,VDM_CPD,RecALS,MHR_CCPD_2,HankMatNN_TC,ESPRIT_MIMO}. While in practice, harmonic components across different modes often exhibit complex mutual correspondences. In MIMO wireless channel data involving ``array-time-frequency'' dimensions, the number of components of the frequency dimension associated with ``clusters'' and the time dimensions associated with ``paths'' \cite{3GPP} are usually different due to resolution differences among the dimensions, resulting in a complex ``one-to-many'' correspondence among harmonic components across the two different modes. In array signal processing using a uniform rectangular array (URA), one single source may arrive at the array from different directions simultaneously, causing the time dimension components of the data to correspond to more than one steering vector in the array dimensions \cite{BTD_ASP}, which are harmonics. Data in these scenarios need to be represented accurately by the BTD, specifically the Rank-$(L_r,L_r,1)$ BTD model.
}

In this paper, we establish a novel MHR model based on multilinear rank (MLR)-$(L_r, L_r, 1)$ block term decomposition (BTD), with Vandermonde structured factor matrices, which efficiently captures the above mentioned ``one-to-many" interaction among MH components, and consider the problem of tensor completion of incomplete observation data based on this model. The problem is transformed into an optimization procedure that aims to fit the BTD model in the least squares (LS) sense, subject to rank minimization of hankelized MH components. By substituting the rank function with the nuclear norm, and introducing auxiliary variables, we finally construct a tractable optimization problem, which can be solved via alternating direction method of multipliers (ADMM).
\ignore{
In this paper, a Rank-$(L_r, L_r, 1)$ BTD based tensor completion method is established for MHR. An optimization problem is constructed by minimizing the rank of hankelized MH components, and the BTD fitting error in the least squares sense. We then formulate a tractable problem and solve it using alternating direction method of multipliers (ADMM), implementing two algorithms for updating the factor matrices. The effectiveness and applicability of our algorithm are demonstrated through both numerical simulations and an application in wireless channel data completion.
}

\textit{Notations:} Scalars, vectors, matrices, and tensors are denoted by italic lowercase, boldface lowercase, boldface uppercase, and calligraphic uppercase letters, respectively. \ignore{We use {\MATLAB} notation to denote subtensors obtained by fixing certain indices or an index range of a tensor.} The transpose is denoted by $(\cdot)^T$. The $\ell_{2}$ norm of vectors, nuclear norm of matrices and Frobenius norm are denoted by $\norm{\cdot}_2,\norm{\cdot}_{\ast}$ and $\norm{\cdot}_{F}$, respectively. We use $\CC^{I_1\!\times\!\cdots\!\times\!I_K}$ to denote the set of tensors of size $I_1\!\times\!\cdots\!\times\!I_K$ with complex values, and $\VV^{J\times L}(\VV^{K})$ to denote the set of Vandermonde matrices of size $J\!\times\! L$ (or vectors of length $K$), respectively.
We use `$\left<\cdot,\cdot\right>$' to denote the real part of the inner product between two matrices.
The outer product is defined as \((\Xcalbf \!\outprod \!\Zcalbf)_{i_1,\cdots,i_M,j_1,\cdots,j_N} \!:=\! (\Xcalbf)_{i_1\!,\!\cdots\!,i_M} (\Zcalbf)_{j_1\!,\!\cdots\!,j_N}\), and the Hadamard product is defined as \((\Pcalbf\!\ast\!\Qcalbf)_{i_1\!,\!\cdots\!,i_M} \!:=\! (\Pcalbf)_{i_1\!,\!\cdots\!,i_M} (\Qcalbf)_{i_1\!,\!\cdots\!,i_M}\).
We use $\rank(\cdot)$ to get the rank of matrices.
\ignore{
For tensor $\Tcalbf\!\in\!\CC^{I\times J\times K}$, we use $\tvec{(\Tcalbf)}:(\tvec{(\Tcalbf)})_{i,j,k}=\Tcalbf_{i+(I-1)j+(IJ-1)k}$ to denote the vector representation of $\Tcalbf$. 
}

\section{Problem Formulation}
\label{sec:formulation}
We consider the following MHR model in which the $(i,j,k)$th entry of a full tensor $\Tcalbf\in\CC^{I\times J\times K}$ can be expressed as the sum of exponential components as follows:
\vspace{-0.5mm}
\begin{equation}\label{eq:origin_entrywise_data_model_cpd}
  t_{i,j,k} \!=\! \sum\nolimits_{r=1}^{R}\left(\sum\nolimits_{l=1}^{L_r} a_{i,r,l}\!\cdot\!\eta_{r,l}^{j-1}\right)\cdot\mu_{r}^{k-1},\; \vspace{-0.5mm}
\end{equation}
where $a_{i,r,l},\eta_{r,l}$ and $\mu_r$ are complex numbers. We define the following vectors:
{\scalefont{1}
\begin{equation}\label{eq:def_vectorset}
\left\{\!\!\begin{aligned}
  \abf_{r,l} &\triangleq [a_{1,r,l},a_{2,r,l},\cdots\!,a_{I,r,l}]^T\in\CC^{I},\\[-2pt]
  \bbf_{r,l} &\triangleq [1,\eta_{r,l},\cdots\!,\eta_{r,l}^{J-1}]^T\in\VV^{J},\\[-2pt]
  \cbf_{r} &\triangleq [1,\mu_{r},\cdots\!,\mu_{r}^{K-1}]^T\in\VV^{K},
\end{aligned}\right.
\end{equation}}%
where $r \!=\! 1,\!\cdots\!,\!R,\, l\!=\!1,\!\cdots\!,\!L_r$.
Note that $\bbf_{r,l}$ and $\cbf_r$ are uniformly sampled complex valued power functions (referred to as harmonics) \footnote{Vector $\abf_{r,l}$ can be either an ordinary vector or a harmonic. As we do not use the structure of this mode in our derivation, we do not consider it a harmonic.}. \ignore{We have $t_{i,j,k}=\sum\nolimits_{r=1}^{R}(\sum\nolimits_{l=1}^{L_r} {\abf}_{r,l}(i)\cdot{\bbf}_{r,l}(j))\cdot{\cbf}_{r}(k)$.} We define $\Tcalbf_{r,l}\!\triangleq\! \abf_{r,l}\outprod\bbf_{r,l}\outprod\cbf_r$ as a third-order rank-1 MH term, and then according to \eqref{eq:origin_entrywise_data_model_cpd}, tensor $\Tcalbf$ can be expressed as the sum of $F\triangleq\sum_{r=1}^{R}L_r$ such terms:
\vspace{-0.5mm}
\begin{equation}\label{eq:def_vdm_btd}
  \Tcalbf = \sum\nolimits_{r=1}^{R}\left( \sum\nolimits_{l=1}^{L_r} \abf_{r,l}\bbf_{r,l}^{T} \right) \outprod \cbf_{r}.\vspace{-0.5mm}
\end{equation}
Note that the $(r,l)$th MH term is generated from the vector group $\{\abf_{r,l}, \bbf_{r,l}, \cbf_r\}$, with $\bbf_{r,l}$ and $\cbf_r$ being harmonics.\linebreak Furthermore, for any fixed index $r$, the index $l$ can vary from $1$ to $L_r$, indicating that a single harmonic $\cbf_r$ in the third mode corresponds to multiple harmonics in the second mode $\{\bbf_{r,1},\!\bbf_{r,2},\!\cdots,\!\bbf_{r,L_r}\}$ belonging to the $r$th part. This establishes a ``one-to-many" correspondence among harmonics across different modes, thereby resulting in an MLR-$(L_r,L_r,1)$ BTD \cite{BTD2} based MHR model.
This model is different from the widely considered CPD based MHR model \cite{HOSVD_MHR,RecALS,MHR_CCPD_2}, in which a simple ``one-to-one'' correspondence is assumed on the harmonics across different modes.

\ignore{
Additionally, when modeling the output data of a URA for far-field narrowband point sources, $F$ represents the number of directions of arrival (DOA) rather than the number of sources. The terms $\boldsymbol{\alpha}_f$, $\boldsymbol{\eta}_f$, and $\boldsymbol{\mu}_f$ denote the source and the steering vectors corresponding to the two array directions associated with the $f$th DOA, respectively. If some sources arrive at the array via multiple paths nearly simultaneously, $\boldsymbol{\alpha}_f$ may be nearly identical for certain values of $f$ \cite{BTD_ASP}.}
\ignore{
Hence, additional constraints should be involved in \eqref{eq:origin_entrywise_data_model_cpd} as follows:
\begin{equation}\label{eq:def_constraint}
  \boldsymbol{\mu}_{1+\sum_{i=1}^{r-1}L_i}=\cdots=\boldsymbol{\mu}_{L_r+\sum_{i=1}^{r-1}L_i},\;\; F = \sum\nolimits_{r=1}^{R} L_r.
\end{equation}
}
\ignore{
This implies that the $F$ components are divided into $R$ parts, with each part containing $L_r$ components. We present the following definitions to differentiate between the models clearly:
\begin{equation}\label{eq:def_btd_fac}
\left\{
  \begin{aligned}
    \abf_{r,l}&\triangleq \boldsymbol{\alpha}_{l+\sum\nolimits_{i=1}^{r-1}L_i},\; &r&=1,\cdots\!,R,\,l=1,\cdots\!,L_r,\\
    \bbf_{r,l}&\triangleq \boldsymbol{\eta}_{l+\sum\nolimits_{i=1}^{r-1}L_i}, &r&=1,\cdots\!,R,\,l=1,\cdots\!,L_r,\\
    \cbf_{r}&\triangleq \boldsymbol{\mu}_{1+\sum\nolimits_{i=1}^{r-1}L_i}, &r&=1,\cdots\!,R.
  \end{aligned}
\right.
\end{equation}

We reformulate model \eqref{eq:origin_entrywise_data_model_cpd} according to \eqref{eq:def_constraint} and \eqref{eq:def_btd_fac} as follows:
\begin{equation}\label{eq:def_vdm_btd}
  \Tcalbf = \sum\nolimits_{r=1}^{R}\left( \sum\nolimits_{l=1}^{L_r} \abf_{r,l}\bbf_{r,l}^{T} \right) \outprod \cbf_{r}.
\end{equation}
Notice that the $(r,l)$th MH is generated from the vector group $\{\abf_{r,l},\bbf_{r,l},\cbf_r\}$ and there is a ``one-to-several" correspondence among harmonics $\bbf_{r,1},\cdots\!,\bbf_{r,L_r}$ and $\cbf_r$. This results in a MLR-$(L_r,L_r,1)$ BTD-based model \cite{BTD2}, which has been studied without considering the inherent components as MHs, but its application in modeling MH signals has not been fully explored.
}
We define the factor matrices of $\Tcalbf$ as follows:
\begin{equation}\label{eq:def_factor_matrix}
  \left\{\begin{aligned}
    \Abf &\triangleq [\Abf_{1},\cdots\!,\Abf_{R}]\in\CC^{I\times F},\,\Abf_{r}\triangleq [\abf_{r,1},\cdots\!,\abf_{r,L_r}],\\[-2pt]
    \Bbf &\triangleq [\Bbf_{1},\cdots\!,\Bbf_{R}]\in\VV^{J\times F},\,\Bbf_{r}\triangleq [\bbf_{r,1},\cdots\!,\bbf_{r,L_r}],\\[-2pt]
    \Cbf &\triangleq [\cbf_{1},\cdots\!,\cbf_{R}]\in\VV^{K\times R}.
  \end{aligned}\right.
\end{equation}
Therefore, it follows from \eqref{eq:def_vdm_btd} that $\Tcalbf$ can be expressed as:
\vspace{-1mm}
\begin{equation}\label{eq:def_vdm_btd_matrix}
  \Tcalbf = \sum\nolimits_{r=1}^{R} \left(\Abf_r\Bbf_r^T\right)\outprod\cbf_r\in\CC^{I\times J\times K}.
\end{equation}
It should be noted that $\Bbf_r$ and $\Cbf$ in \eqref{eq:def_factor_matrix}, which hold the harmonics $\bbf_{r,l},\cbf_{r}$ as columns, are Vandermonde matrices.
\ignore{
Besides, for an unconstrained MLR-$(L_r,L_r,1)$ BTD, the factor matrix $\Cbf$ is essentially unique under mild conditions with scaling and permutation ambiguities, but $\Abf$ and $\Bbf$ are unique with not only scaling and permutation ambiguities but also rotation ambiguities \cite{BTD2}. However, {\hl{as will be shown later}}, the harmonic structure may lead to the uniqueness of $\Abf$ and $\Bbf$ with no rotation ambiguity.
}

The model \eqref{eq:def_vdm_btd} is frequently encountered in practice. Taking the aforementioned wireless Sub-6GHz CSI model as an example, according to the standard 3GPP communication protocol \cite{3GPP}, the wireless CSI in ``space-time-frequency" domain under the Sub-6GHz spectrum can be expressed as:
\vspace{-2pt} 
\begin{equation}\label{eq:def_channel_model}
\Ccalbf = \sum\nolimits_{r=1}^{R}\left(\sum\nolimits_{l=1}^{L_r}\abf_{r,l}\bbf_{r,l}^{T}\right)\outprod \cbf_r,
\end{equation}
where $R$ and $L_r$ are the number of ``clusters'' and the number of ``paths'' in the $r$th ``cluster'', respectively. The vectors $\abf_{r,l}$ and $\bbf_{r,l}$ are the array steering vectors and the harmonics in the time domain respectively, which are  associated with the $l$th ``path'' in the $r$th ``cluster'', $\cbf_r$ is the harmonic in the frequency domain associated with the $r$th ``cluster''. This leads to the BTD based MHR model in \eqref{eq:def_vdm_btd}.

In this paper, we aim to recover the full $\Tcalbf$, which is an MH signal, from a small subset of its observed entries. We define a binary index tensor $\mathcalbf{W}$ of size $I\times J\times K$, in which the $(i,j,k)$th entry $w_{\!i,j,k}$ is $1$ if $t_{i,j,k}$ is observed and is $0$ otherwise. Hence, the sampled tensor can be represented by $\mathcalbf{Y}\!\triangleq\!\mathcalbf{W}\!\ast\!\Tcalbf$, where the missing values are filled with $0$. Therefore, our MHR completion problem can be stated as follows: with observed incomplete data tensor $\Ycalbf$ and binary index tensor $\Wcalbf$, find factor matrices $\Abf,\Bbf$, and $\Cbf$ such that:

{(1)} the data tensor reconstructed by estimates of of $\Abf,\Bbf,\Cbf$ via MLR-$(L_r,L_r,1)$ BTD model best fit $\Ycalbf$ at partially observed entries in the LS sense;

{(2)} factor matrices $\Bbf$ and $\Cbf$ are Vandermonde matrices.

Mathematically, the above problem can be expressed as:
\vspace{-0.2mm}
\begin{equation}\label{eq:def_vdm_btd_withmissing}
\left\{\begin{aligned}
  &\min_{\Abf,\Bbf,\Cbf}\;\,f_1\!\triangleq\!\norm{\Ycalbf\!-\!\mathcalbf{W}\!\ast\!\left(\sum\nolimits_{r=1}^{R}\! \left(\Abf_r\Bbf_r^T\right)\outprod\cbf_{r}\right)}_{F}^{2},\\[-0.7mm]
  &\;\;\,\mathrm{s.t.,}\;\;\;\Bbf,\Cbf \text{ are Vandermonde matrices.}\quad\quad\quad\quad
\end{aligned}\right.
\end{equation}

\ignore{
\begin{equation}\label{eq:def_vdm_btd_withmissing}
\left\{\begin{aligned}
  &\mathcalbf{W}\ast\Tcalbf = \mathcalbf{W}\ast \left(\sum\nolimits_{r=1}^{R}\! \left(\Abf_r\Bbf_r^T\right)\outprod\cbf_{r}\right) \text{ is known},\\
  &\text{Find } \Abf,\Bbf,\Cbf: \;\Abf\!\in\!\CC^{I\times F},\Bbf\!\in\!\VV^{J\times F},\Cbf\!\in\!\VV^{K\times R}.
\end{aligned}\right.
\end{equation}
}
\ignore{
We note that although algorithms based on CPD can also be applied to solve \eqref{eq:def_vdm_btd_withmissing}, they tend to increase the number of potential parameters, raise the complexity, and potentially compromise the performance.
Besides, though the proposed algorithm is derived from the tensors used to model the wireless channel data, it can be easily extended to tensors that model data in other practical scenes and other higher-order tensors composed of MH components.
}

\section{Proposed Algorithm}
\label{sec:algorithm}
\ignore{
The problem \eqref{eq:def_vdm_btd_withmissing} requires the decomposition of an incomplete BTD tensor, necessitating the use of optimization methods. However, the non-convex feasible lead by the Vandermonde structure makes the problem difficult to solve. In the following, we transform it into a tractable optimization problem with equality constraints \cite{HankMatNN_TC}, and propose an ADMM based solver for this problem. Then, the full tensor $\Tcalbf$ can be recovered via the estimated \({\mathbf{A}},{\mathbf{B}},{\mathbf{C}}\). The steps are as follows:
}
\ignore{
This tensor completion problem \eqref{eq:def_vdm_btd_withmissing} necessitates the use of optimization methods. However, the Vandermonde structure leads to a specially feasible region and thereby a difficult optimization problem. In the following, we transform it into a tractable optimization problem with equality constraints \cite{HankMatNN_TC}, and propose an ADMM based solver for this problem. Then, the full tensor $\Tcalbf$ can be recovered via the estimated \({\mathbf{A}},{\mathbf{B}},{\mathbf{C}}\). \ignore{The steps are as follows:}
}
\ignore{
In this section, we propose an optimization-based tensor completion algorithm, which jointly exploits the MLR-$(L_r, L_r, 1)$ BTD structure of the data and the Vandermonde structure of matrices $\Bbf$ and $\Cbf$, to solve the problem in \eqref{eq:def_vdm_btd_withmissing}.
}

\subsection{Transformation into a Tractable Problem}
We note that a harmonic (or a column of a Vandermonde matrix) has the property that its hankelized matrix is rank-1 \cite{Hankel_Perform},\linebreak and thus enforcing a matrix to be Vandermonde can be treated as minimizing the sum of ranks of its hankelized columns:
\vspace{-0.5mm}
\begin{equation}\label{eq:def_Loss_harmonic}
  \min_{\Bbf,\Cbf}\; \sum_{r=1}^{R}\!\sum_{l=1}^{L_r} \rank(\Hcal(\bbf_{r,l}))+\sum_{r=1}^{R}\rank(\Hcal(\cbf_{r})),
  \vspace{-0.3mm}
\end{equation}
where $\Hcal(\cdot)$ denotes the hankelization operator \cite{Hankel_Perform}. Noting that the above rank optimization problem is non-convex, we use a common convex relaxation approach that replaces the rank functions in \eqref{eq:def_Loss_harmonic} by nuclear norms, the minimization of which, under mild conditions \cite{SVT,Hankel_Perform}, can achieve the goal\pagebreak $\;$of rank minimization. Therefore, the problem \eqref{eq:def_Loss_harmonic} can be further relaxed into a new problem as follows:
\vspace{-1mm}
\begin{equation}\label{eq:def_Loss_harmonic_eq}
  \min_{\Abf,\Bbf,\Cbf}\; f_2 \!\triangleq\! \sum_{r=1}^{R}\!\sum_{l=1}^{L_r} \!\norm{\Hcal(\bbf_{r,l})}_{\!\ast}\!+\!\sum_{r=1}^{R}\!\norm{\Hcal(\cbf_{r})}_{\!\ast}\!+\!\norm{\Abf}_{F}^2.
\end{equation}
where the squared Frobenius norm of $\Abf$ is introduced to balance the level of the factor matrices. Hence, problem 
\eqref{eq:def_vdm_btd_withmissing} can be transformed into a multi-objective optimization problem:
\begin{equation}\label{eq:def_mutiobj_opt}
  \min\nolimits_{\Abf,\Bbf,\Cbf}\;\, \fbf(\Abf,\!\Bbf,\!\Cbf) \!\triangleq\! [f_1(\Abf,\!\Bbf,\!\Cbf),f_2(\Abf,\!\Bbf,\!\Cbf)]^T,
\end{equation}
where $f_1$ is the MHR-$(L_r, L_r, 1)$ BTD model fit error in the LS sense defined in \eqref{eq:def_vdm_btd_withmissing}, and $f_2$ is the sum of nuclear norms of hankelized harmonics defined in \eqref{eq:def_Loss_harmonic}. Employing the linear weighted sum method, the above \ignore{multi-objective}optimization problem \eqref{eq:def_mutiobj_opt} can be further transformed into the following problem:
\begin{equation}\label{eq:lws_method}
  \min\nolimits_{\Abf,\Bbf,\Cbf}\; \lambda f_1(\Abf,\Bbf,\Cbf)+f_2(\Abf,\Bbf,\Cbf),
\end{equation}%
where \(\lambda\) balances the two objectives. Normally, we set \(\lambda\) based on prior knowledge on the overall mulinear data model and the harmonic structure of factor matrices. For instance, in low signal-to-noise rate (SNR), the model error is expected to be large and we tend to set \(\lambda\) small to emphasize the latter term $f_2$. {{Besides, completion can be performed for various $\lambda$ values, with the best result chosen based on the degree to which columns of the estimated factor matrices $\smash{\hat{\Bbf},\hat{\Cbf}}$ are harmonics. A strategy based on this with known SNR will be detailed in the extended version due to space limits.}}

\ignore{We introduce auxiliary 
This problem involves two aspects: (1) the BTD constraint on the tensor, and (2) the Vandermonde constraints on the factor matrices $\Bbf_r,\Cbf$. We address these constraints by formulating two separate optimization problems, which are then combined into a multi-objective optimization framework. Specifically,
\ignore{
\begin{itemize}[left=0pt, noitemsep, topsep=0pt, label=--]
  \item BTD constraint:
  We note that it can be easily realized by minimizing the model fitting error in the least squares sense as follows:
  \begin{equation}\label{eq:def_Loss_btd}
    \min_{\Abf,\Bbf,\Cbf}\,f_1 \triangleq \norm{\Ycalbf\!-\!\mathcalbf{W}\ast\!\sum\nolimits_{r=1}^{R}\! \left(\Abf_r\Bbf_r^T\right)\outprod\cbf_{r}}_{F}^2.
  \end{equation}
  \item Vandermonde constraints: 
\end{itemize}
}

\textit{(1) BTD constraint:} This can be easily achieved by minimizing the fitting error in the least squares sense as follows:
\begin{equation}\label{eq:def_Loss_btd}
\min_{\Abf,\Bbf,\Cbf}\,f_1 \triangleq \norm{\Ycalbf\!-\!\mathcalbf{W}\ast\!\sum\nolimits_{r=1}^{R}\! \left(\Abf_r\Bbf_r^T\right)\outprod\cbf_{r}}_{F}^2.
\end{equation}
\ignore{
\textit{(2) Vandermonde constraints:} This is comparatively more challenging, as quantification methods are needed for assessing the extent to which $\Bbf_r,\Cbf$ are Vandermonde. Since hankelized harmonics yield rank-1 matrices \cite{Hankel_Perform}, we aim to minimize the rank of the hankelized MH components as follows:
}

\textit{(2) Vandermonde constraints:} This is comparatively more challenging, as quantification methods are required to assess the extent to which $\Bbf_r$ and $\Cbf$ exhibit Vandermonde structure. Given that hankelized harmonics are rank-1 matrices \cite{Hankel_Perform}, we aim to minimize the rank of the hankelized MH components:
\begin{equation}\label{eq:def_Loss_harmonic}
  \min_{\Abf,\Bbf,\Cbf}\; \sum_{r=1}^{R}\!\sum_{l=1}^{L_r} \rank(\Hcal(\bbf_{r,l}))+\sum_{r=1}^{R}\rank(\Hcal(\cbf_{r})),
\end{equation}
where \(\mathcal{H}(\cdot)\) is the hankelization operator. However, the rank optimization problem is non-convex, and a popular convex relaxation is based on minimization of the nuclear norm under certain soft conditions \cite{SVT,Hankel_Perform}. We follow this method and relax \eqref{eq:def_Loss_harmonic} to a new problem as follows:
\begin{equation}\label{eq:def_Loss_harmonic_eq}
  \min_{\Abf,\Bbf,\Cbf}\; f_2 \!\triangleq\! \sum_{r=1}^{R}\!\sum_{l=1}^{L_r} \!\norm{\Hcal(\bbf_{r,l})}_{\!\ast}\!+\!\sum_{r=1}^{R}\!\norm{\Hcal(\cbf_{r})}_{\!\ast}\!+\!\norm{\Abf}_{F}^2.
\end{equation}
where the squared Frobenius norm of $\Abf$ is introduced to balance the scale of the factor matrices.

Therefore, the problem \eqref{eq:def_Loss_harmonic_eq} can be transformed into a multi-objective optimization problem as follows:
\begin{equation}\label{eq:def_mutiobj_opt}
  \min\nolimits_{\Abf,\Bbf,\Cbf}\;\, \fbf(\Abf,\!\Bbf,\!\Cbf) \!\triangleq\! [f_1(\Abf,\!\Bbf,\!\Cbf),f_2(\Abf,\!\Bbf,\!\Cbf)]^T,
\end{equation}

Hence, we can in fact transform problem \eqref{eq:def_vdm_btd_withmissing} into a multi-objective optimization problem by combining \eqref{eq:def_Loss_btd} and \eqref{eq:def_Loss_harmonic_eq}:
\begin{equation}\label{eq:def_mutiobj_opt}
  \min\nolimits_{\Abf,\Bbf,\Cbf}\;\, \fbf(\Abf,\!\Bbf,\!\Cbf) \!\triangleq\! [f_1(\Abf,\!\Bbf,\!\Cbf),f_2(\Abf,\!\Bbf,\!\Cbf)]^T,
\end{equation}

In this paper, we employ the linear weighted sum method to solve it, which means we transform it as follows:
\begin{equation}\label{eq:lws_method}
  \min\nolimits_{\Abf,\Bbf,\Cbf}\; \lambda f_1(\Abf,\Bbf,\Cbf)+f_2(\Abf,\Bbf,\Cbf),
\end{equation}%
where $\lambda$ is a parameter that balances the harmonic structure of the underlying factors and the BTD structure of the data, which should be set to a smaller value under high noise to emphasize more on the data.}We introduce auxiliary variables for managing the hankelization operator in $\norm{\cdot}_{\ast}$ and finally transform the problem \eqref{eq:lws_method} into the following tractable optimization problem:
\begin{equation}\label{eq:aux_vars}
\left\{\!\!\!\begin{aligned}
    &\min_{\underset{\{\!\Ebf_{r\!,l}\!\}\!,\!\{\!\Fbf_{\!r}\!\}}{\Abf,\Bbf,\Cbf}}f\!\triangleq\! \lambda f_1\!+\!\!\sum_{r=1}^{R}\!\sum_{l=1}^{L_r} \!\norm{\Ebf_{r\!,l}}_{\!\ast} \!+\!\! \sum_{r=1}^{R} \!\norm{\Fbf_{\!r}}_{\!\ast}\!\!+\!\norm{\Abf}_{F}^2,\\[-4pt]
    &\mathrm{\quad\, s.t.,\;\;}\Ebf_{r,l}= \Hcal(\bbf_{r,l}),\;\Fbf_r=\Hcal(\cbf_{r}).
\end{aligned}\right.
\end{equation}%
\subsection{ADMM-based Method}
We solve \eqref{eq:aux_vars} using ADMM, which handles the constraint via the augmented Lagrangian method, while alternates the optimization among variable groups and updates the Lagrange multipliers. The augmented Lagrangian function of \eqref{eq:aux_vars} is given as follows:
\vspace{-2mm}
\begin{equation*}\label{eq:final_cost_fun}
{\scalefont{1}
\begin{aligned}
  &f_{\text{Lag}}\!\triangleq\!\lambda f_1\!+\! \norm{\Abf}_F^{2}\!+\!\sum_{r=1}^{R}\!\left< \Nbf_{r},\!\Hcal(\cbf_{r})\!-\!\Fbf_{r} \right> \!+\! \beta\!\norm{\Hcal(\cbf_{r})\!-\!\Fbf_{r}}_{F}^2\\[-5pt]
  &+\sum_{r=1}^{R}\!\sum_{l=1}^{L_r}\!\left< \Mbf_{r,l},\!\Hcal(\bbf_{r,l})\!-\!\Ebf_{r,l} \right> \!+\! \beta\!\norm{\Hcal(\bbf_{r,l})\!-\!\Ebf_{r,l}}_{F}^2,
\end{aligned}}
\end{equation*}%
where $\Mbf_{r,l}\in\CC^{J_1\times J_2},\Nbf\in\CC^{K_1\times K_2}$ are Lagrange multipliers for $r=1,\cdots\!,R,l=1,\cdots\!,L_r$,
$\beta$ is the penalty parameter with a small initial value.
\ignore{
The variables are divided into three groups: $\{\Abf,\Bbf,\Cbf\}$, $\{\Ebf_{r, l},\Fbf_{r}\}$ and $\{\Mbf_{r,l},\Nbf_{r}\}$. we update the variable groups sequentially at each iteration.}Specifically, the following steps are performed sequentially at the $(k+1)$th iteration:

\ignore{
\begin{itemize}[label=\textit{(\arabic*)},left=0pt,itemsep=\parskip, parsep=0pt,topsep=-2pt,labelindent=1em]
  \item Updating the $\{\Abf,\Bbf,\Cbf\}$ by minimizing $\Lcal$;
  
  \item For every $r,l$, updating $\{\Ebf_{r,l},\Fbf_{r}\}$ by minimizing $\Lcal$ using the singular value thresholding (SVT) algorithm \cite{SVT};
      
  \item For every $r,l$, updating $\{\Mbf_{r,l},\!\Nbf_{r}\!\}$ by the gradient ascent method, with a convergence-ensuring step size of $\beta$ \cite{HankMatNN_TC};
      
  \item Updating the penalty parameter $\beta^{(k+1)}=\rho\beta^{(k)}$ to accelerate the convergence \cite{HankMatNN_TC}, where $\rho\in(1.0,1.1]$.%
\end{itemize}
\begin{itemize}[left=0pt,itemsep=\parskip, parsep=0pt,topsep=-2pt,labelindent=0em,leftmargin=0pt,listparindent=2em]
  \item Updating the $\{\Abf,\Bbf,\Cbf\}$ by minimizing $\Lcal$;
  
  \item For every $r,l$, updating $\{\Ebf_{r,l},\Fbf_{r}\}$ by minimizing $\Lcal$ using the singular value thresholding (SVT) algorithm \cite{SVT};
      
  \item For every $r,l$, updating $\{\Mbf_{r,l},\!\Nbf_{r}\!\}$ by the gradient ascent method, with a convergence-ensuring step size of $\beta$ \cite{HankMatNN_TC};
      
  \item Updating the penalty parameter $\beta^{(k+1)}=\rho\beta^{(k)}$ to accelerate the convergence \cite{HankMatNN_TC}, where $\rho\in(1.0,1.1]$.%
\end{itemize}
}

\par{\makebox[5.5mm][l]{(a)}}Update factor matrices $\{\Abf,\!\Bbf,\!\Cbf\}$ by minimizing $f_{\text{Lag}}$;
\par{\makebox[5.5mm][l]{(b)}}Update $\{\Ebf_{r,l},\Fbf_{r}\}$ for every $r,l$, by minimizing $f_{\text{Lag}}$ using the singular value thresholding (SVT) algorithm \cite{SVT};\pagebreak
\par{\makebox[5.5mm][l]{(c)}}Update $\{\Mbf_{r,l},\Nbf_{r}\}$ for every $r,l$, by the gradient ascent method, with a convergence-ensuring step size of $\beta$ \cite{HankMatNN_TC};\linebreak
\vspace{-4mm}
\par{\makebox[5.5mm][l]{(d)}}Update the penalty parameter $\beta^{(k+1)}=\rho\beta^{(k)}$ to accelerate the convergence \cite{HankMatNN_TC}, where $\rho\in(1.0,1.1]$.%
\ignore{
\textit{Remark 1:} The subproblems mentioned in \textit{(2)} are solved using the Singular Value Thresholding (SVT) algorithm \cite{SVT}. This operation shrinks the small singular values of the Hankel matrices and implicitly promotes the low-rank structure.
}

Since the sub-problems in steps (b)--(d) already admit solutions from existing literature, we herein only address the sub-problem in step (a). More specifically, two methods for solving this problem will be given in Subsection 3.3.

\vspace{-0.7mm}
\subsection{Factor Matrices Updating}
We re-write the sub-problem in step (a) as follows:
\vspace{-1mm}
\begin{equation}\label{eq:related_cost_fun}
\begin{aligned}
   \min_{\Abf\!,\Bbf\!,\Cbf}\;\,&g(\Abf,\Bbf,\Cbf)\triangleq\lambda%
   \ignore{\norm{\Pcal_\Omega(\Tcalbf)\!-\!\Pcal_\Omega\!\left(\sum_{r=1}^{R}\! \sum_{l=1}^{L_r} \abf_{r,l}\bbf_{r,l}^{T}\outprod\cbf_{r}\!\right)\!}_{F}^2\\[-7pt]}%
   f_1(\Abf,\Bbf,\Cbf)+\norm{\Abf}_{F}^2\\[-7pt]
   +&\beta\sum\nolimits_{r=1}^{R}\!\sum\nolimits_{l=1}^{L_r}\!\norm{\Hcal(\bbf_{r,l})\!-\!\Ebf_{r,l}^{(k)}\!+\!\beta^{-1}\Mbf_{r,l}^{(k)}}_{F}^{2}\\
   +&\beta\sum\nolimits_{r=1}^{R}\norm{\Hcal(\cbf_r)\!-\!\Fbf_r^{(k)}\!+\!\beta^{-1}\Nbf_r^{(k)}}_{F}^{2},
\end{aligned}
\vspace{-1mm}
\end{equation}
where the superscript `$(k)$' of a variable denotes its value after the $k$th iteration. This is a LS problem, which can be solved using iterative algorithms. Two algorithms are considered:

{\makebox[5.5mm][l]{\textit{(1)} }}\textit{Alternating LS (ALS)}

We minimize the objective function of \eqref{eq:related_cost_fun} in the $(n\!+\!1)$th iteration as follows: (a1) update $\Abf$ by solving \eqref{eq:key_to_ALS_A}, with $\Bbf$ and $\Cbf$ fixed; (a2) update $\Bbf$ by solving \eqref{eq:key_to_ALS_B}, with $\Abf$ and $\Cbf$ fixed; (a3) update $\Cbf$ by solving \eqref{eq:key_to_ALS_C}, with $\Abf$ and $\Bbf$ fixed.
\vspace{-0.2mm}
\begin{align}
  \Abf^{(n+1)} &= \argmin_{\Abf}\;\, g(\Abf,\Bbf^{(n)},\Cbf^{(n)}),\label{eq:key_to_ALS_A}\\[-1pt]
  \Bbf^{(n+1)} &= \argmin_{\Bbf}\;\, g(\Abf^{(n+1)},\Bbf,\Cbf^{(n)}),\label{eq:key_to_ALS_B}\\[-1pt]
  \Cbf^{(n+1)} &= \argmin_{\Cbf}\;\, g(\Abf^{(n+1)},\Bbf^{(n+1)},\Cbf)\label{eq:key_to_ALS_C}.
\end{align}
Readers may refer to the appendix of \cite{HankMatNN_TC} for further derivations.
\ignore{ Taking $\Bbf^{(k)}$ as an example, we calculate the gradient of the cost function with respect to $\Bbf$ and set the gradient to $\mathbf{0}$:
\begin{equation}\label{eq:als_upd_B}
  \Bbf\left(\lambda \Gbf^{(\Bbf)} - \right)
\end{equation}}%

{\makebox[5.5mm][l]{\textit{(2)} }}\textit{Gauss-Newton Nonlinear LS (GN-NLS)}

\ignore{Let $\zbf\!\triangleq\![\tvec(\Abf)^{T}\!,\!\tvec(\Bbf)^{T}\!,\!\tvec(\Cbf)^{T}]^{T}$ denote the optimization variables.}
Let $\zbf$ denote the optimization variable that collects all the entries of $\Abf,\Bbf,\Cbf$.
We note that the term $f_1$, which is the Frobenius norm of the vectorized residual tensor $\mathbf{r}(\zbf)$, leads to a nonlinear LS problem. At the $(n\!+\!1)$th iteration, GN-NLS approximate $\mathbf{r}(\zbf)$ using its first-order Taylor expansion around $\zbf^{(n)}$ as $\mathbf{r}(\zbf)\!\approx\!\rbf(\zbf^{(n)})+\Jbf(\zbf^{(n)})\pbf$, with $\Jbf(\zbf)$ being the Jacobian of $\rbf(\zbf)$ and $\pbf\triangleq\zbf-\zbf^{(n)}$ being the step. Hence, $\zbf$ is updated via an optimal step $\pbf^{\star}$ obtained by solving an optimization problem analogous to linear LS as follows:
\begin{equation}\label{eq:eq:approximate_related_cost_fun}
\min\nolimits_{\pbf}\;g\approx\lambda|\kern-0.1em|\rbf(\zbf^{(n)})+\Jbf(\zbf^{(n)})\pbf|\kern-0.1em|_2^2 + h(\pbf+\zbf^{(n)}).
\end{equation}
where \( h(\cdot) \) denote the terms in \( g \) excluding \( f_1 \), consisting of standard second-order terms as hankelization is linear.
The dogleg trust region method is used by us to compute \(\pbf^{\star}\)\ignore{, which combines the steepest descent and the GN method}\ignore{., ensuring the step remains within the trust region with a radius \(\Delta^{(n)}\)}\ignore{, i.e., $\norm{\pbf}_{2}\!\leq\! \Delta^{\!(n)}\!$}. We note that $\Jbf(\zbf)$ is sparse for both the BTD structure \cite{CPD_NLS} and the observation operation \cite{tomasi2005parafac}, which is used to simplify calculations. Readers may refer to \cite{CPD_NLS,tomasi2005parafac} for further derivations.

Besides, it is noted that finding the exact optimal solution for a single sub-problem, possibly via a number of iterations, is unnecessary. Therefore, we choose to iterate only once to obtain a reasonably accurate one.

\ignore{
\vspace{-2mm}
\subsection{Optimal Selection from Completion Results}
The parameter \(\lambda\) is crucial for tensor completion. Given the SNR defined from \eqref{eq:constructed_sampled_tensor}, we complete the tensor for various values of $\lambda$ in a predefined range \(\mathbb{L}\), and select a result based on the criteria:
(a) The hankelized harmonics are nearly rank-1;
(b) The residual-to-signal ratio (RSR) should be close to the theory RSR informed by the known SNR.
Let \(\hat{\mathbf{A}}(\lambda)\), \(\hat{\mathbf{B}}(\lambda)\), \(\hat{\mathbf{C}}(\lambda)\) and \(\hat{\boldsymbol{\mathcal{T}}}(\lambda)\) denote the estimated factor matrices and the completed tensor corresponding to \(\lambda\), respectively.
Specially,
{\scalefont{1}
\begin{equation*}\label{eq:opt_lambda}
  \lambda^{\star}\!=\!\underset{\lambda\in\LL^{\prime}}{\arg\max} \sum_{r=1}^{R}\!\sum_{l=1}^{L_r} \!\frac{|\kern-1.5pt|\Hcal(\hat{\bbf}_{r,l}(\lambda))|\kern-1.5pt|_2}{|\kern-1.5pt|\Hcal(\hat{\bbf}_{r,l}(\lambda))|\kern-1.5pt|_F}\! + \!\sum_{r=1}^{R}\!\frac{|\kern-1.5pt|\Hcal(\hat{\cbf}_{r}(\lambda))|\kern-1.5pt|_2}{|\kern-1.5pt|\Hcal(\hat{\cbf}_{r}(\lambda))|\kern-1.5pt|_F},
\end{equation*}}%
where $\LL^{\prime}$ is a subset of $\LL$ defined as follows:
{\scalefont{1}
\begin{equation*}\label{eq:def_LLrange}
  \LL^{\prime}\!=\{\lambda\!\in\!\LL\left|\right.\! \gamma_\text{L}\cdot s < |\!|\Ycalbf|\!|_{\!F}^{-2}\cdot|\!|\Wcalbf\ast\hat{\Tcalbf}(\lambda)\!-\!\Ycalbf|\!|_{\!F}^2< \gamma_\text{H}\cdot s\},
\end{equation*}}%
where $s\!\triangleq\!10^{\mathrm{SNR}/{10}} \!+\! 1$ is the theory SRS, $\gamma_{\text{L}},\gamma_{\text{H}}$ are tuning parameters not large than $1$ to allow the overfitting to noise. The we select $\hat{\Tcalbf}(\lambda^{\star})$ as the completed tensor.
\ignore{
The parameter \(\lambda\) plays a pivotal role in tensor completion. To find the optimal \(\lambda\), we compute the completed tensor for various values in a predefined range \(\mathbb{L}\), evaluating the results based on specific criteria. Given the known signal-to-noise ratio (SNR) derived from \eqref{eq:constructed_sampled_tensor}, the factor matrices corresponding to each \(\lambda\) are denoted by \(\hat{\mathbf{A}}(\lambda)\), \(\hat{\mathbf{B}}(\lambda)\), \(\hat{\mathbf{C}}(\lambda)\), while the reconstructed full tensor is represented by \(\hat{\boldsymbol{\mathcal{T}}}(\lambda)\).

The selection of the optimal \(\lambda\) is driven by two considerations:
(a) The hankelized harmonics of the reconstructed tensor should be nearly rank-1.
(b) The residual-to-signal ratio (RSR) should be slightly smaller, yet close to the approximate ratio informed by the known SNR.
}%
\ignore{
This approach balances the rank constraint on the harmonics with a reasonable fit to the original tensor's noise level, ensuring that the completion result is both accurate and adheres to the expected signal characteristics.

The parameter $\lambda$ is crucial for tensor completion. We complete the tensor for various $\lambda$ values in a range $\mathbb{L}$ and select the best result. Given the known SNR from \eqref{eq:constructed_sampled_tensor}, the estimated factor matrices are denoted as $\hat{\Abf}(\lambda), \hat{\Bbf}(\lambda), \hat{\Cbf}(\lambda)$ and the full tensor as $\hat{\Tcalbf}(\lambda)$. The optimal $\lambda$ is chosen to make the hankelized harmonics nearly rank-1, while ensuring the ``residual-to-signal ratio" (RSR) stays slightly smaller but close to the approximate ratio based on the known SNR. Specially,
}%

\ignore{
The parameter $\lambda$ is crucial for tensor completion. We complete the tensor for various $\lambda$ values in a range $\mathbb{L}$ and select the best result. Given the known SNR from \eqref{eq:constructed_sampled_tensor}, the estimated factor matrices are denoted as $\hat{\Abf}(\lambda), \hat{\Bbf}(\lambda), \hat{\Cbf}(\lambda)$ and the full tensor as $\hat{\Tcalbf}(\lambda)$. The optimal $\lambda$ is chosen to make the hankelized harmonics nearly rank-1, while ensuring the ``residual-to-signal ratio" (RSR) stays slightly smaller but close to the approximate ratio based on the known SNR. Specially,
Specially,
{\scalefont{1}
\begin{equation*}\label{eq:opt_lambda}
  \lambda^{\star}\!=\!\underset{\lambda\in\LL^{\prime}}{\arg\max} \sum_{r=1}^{R}\!\sum_{l=1}^{L_r} \!\frac{|\kern-1.5pt|\Hcal(\hat{\bbf}_{r,l}(\lambda))|\kern-1.5pt|_2}{|\kern-1.5pt|\Hcal(\hat{\bbf}_{r,l}(\lambda))|\kern-1.5pt|_F}\! + \!\sum_{r=1}^{R}\!\frac{|\kern-1.5pt|\Hcal(\hat{\cbf}_{r}(\lambda))|\kern-1.5pt|_2}{|\kern-1.5pt|\Hcal(\hat{\cbf}_{r}(\lambda))|\kern-1.5pt|_F},
\end{equation*}}%
where $\LL^{\prime}$ is a subset of $\LL$ obtained by removing the $\lambda$ valus that cause RSR to be overlarge or oversmall, defined as follows:
{\scalefont{1}
\begin{equation*}\label{eq:def_LLrange}
  \LL^{\prime}\!=\left\{\!\lambda\!\in\!\LL\left|\right.\! \frac{\gamma_\text{L}}{10^{\frac{\mathrm{SNR}}{10}} \!+\! 1} \!<\! \frac{\!|\!|\Wcalbf\ast\hat{\Tcalbf}(\lambda)\!-\!\Ycalbf|\!|_{\!F}^2}{{|\!|\Ycalbf|\!|_{\!F}^2}}\!<\! \frac{\gamma_\text{H}}{10^{\frac{\mathrm{SNR}}{10}} \!+\! 1}\right\},
\end{equation*}}%
where $\gamma_{\text{L}},\gamma_{\text{H}}$ are tuning parameters, where both not large than $1$ to allow the overfitting to noise. 
}
}
\ignore{
We use $\zbf\!\triangleq\![\tvec(\Abf)^{T}\!,\!\tvec(\Bbf)^{T}\!,\!\tvec(\Cbf)^{T}]^{T}$ to denote the optimization variables and approximate the nonlinear term $f_1$ in $g$ using the first-order Taylor expansion of the residual $\pbf(\zbf)\!\triangleq\!\tvec(\mathcalbf{Y}\!-\!\mathcalbf{W}\ast\sum_{r=1}^{R}\! (\Abf_{r}\Bbf_{r}^{T})\outprod\cbf_{r})$, where The Jacobian defined as $\Jbf(\zbf)\!=\!\mathrm{d}\pbf(\zbf) / \mathrm{d}\zbf$, is sparse for two-fold reasons: the inherent ``one-to-many" BTD structure \cite{CPD_NLS} and the observation \cite{tomasi2005parafac}. We focus only on the non-sparse parts in the calculation. The readers can refer to \cite{CPD_NLS,tomasi2005parafac} for further derivations.
}

\section{Experimental Results}
\label{sec:experimental}
We present experimental results to demonstrate the performance of the proposed algorithms, termed as MHR-TC-BTD-ALS(NLS). The compared tensor completion methods include the CPD based method for MHR \cite{HankMatNN_TC} (MHR-TC-CPD), the CPD based method using NLS \cite{tomasi2005parafac} (TC-CPD-NLS), \ignore{the CPD based tensor completion method}using first-order optimization \cite{ACAR201141} (TC-CPD-WOPT), and the BTD based method using reweighted LS recursions with a hierarchical structure \cite{BTDCOM_rank} (TC-BTD-HIRLS). Note that the last three methods do not make use of the harmonic structure\ignore{ in the completion}\ignore{, while the proposed methods and MHR-TC-CPD have taken into account both the harmonic\ignore{ structure} and the multilinear structure (CPD or BTD)}.

\vspace{-0.5mm}
\subsection{Numerical Simulations}
In the experiments, \( \eta_{r,l} \) and \( \mu_{r} \) are independently and uniformly sampled from the unit circle in the complex plane, and the entries of $\Abf$ are drawn independently from a standard complex Gaussian distribution and the noiseless full tensor $\Tcalbf^\prime \in \CC^{I \times J \times K}$ is generated according to \eqref{eq:def_vdm_btd}. The noisy tensor to be observed is then constructed as follows:
\vspace{-1mm}
\begin{equation}\label{eq:constructed_sampled_tensor}
  \Tcalbf =  \Tcalbf^\prime + (\sigma_n \cdot \norm{\Tcalbf^\prime}_F) / (\sigma_s \cdot \norm{\Ncalbf}_F) \cdot \Ncalbf,
  \vspace{-1mm}
\end{equation}
where $\Ncalbf$ is a noise tensor, with entries drawn from a standard complex Gaussian distribution independently. The SNR is defined as $\text{SNR}=20\log_{10}(\sigma_s/\sigma_n) \text{ (dB)}$. Besides, each entry of $\Tcalbf$ is equiprobably observed and the proportion of sampled entries is denoted by $\rho\triangleq\sum_{i,j,k}{w_{i,j,k}}/(IJK)$.

The relative least normalized error (RLNE) \cite{HankMatNN_TC} is used to evaluate the performance of algorithms, which is defined as:
\vspace{-1mm}
\begin{equation}\label{eq:def_RFE}
  \text{RLNE} =\min \{\|\hat{\Tcalbf} - \Tcalbf^\prime\|_{F} / \norm{\Tcalbf^\prime}_{F},1\},
  \vspace{-2mm}
\end{equation}
where $\hat{\Tcalbf}$ is the estimated completed tensor of each algorithm. \ignore{We use $\hat{\Abf}, \hat{\Bbf}, \hat{\Cbf}$ to\linebreak denote the estimated factor matrices, then $\hat{\Tcalbf}$ is calculated as $\hat{\Tcalbf} = \sum_{r=1}^{R} \sum\nolimits_{l=1}^{L} (\hat{\Abf}_{r} \hat{\Bbf}_r^{T})\circ \hat{\cbf}_r$ for the BTD based algorithms and $\hat{\Tcalbf} = \sum_{f=1}^{F} \hat{\abf}_{f}\outprod\hat{\bbf}_f\outprod\hat{\cbf}_f$ for the CPD based algorithms.}

\begin{figure}[htb]
\vspace{-1.7mm}
\begin{minipage}[b]{0.495\columnwidth}
  \centering
  \centerline{\hspace{3mm}\includegraphics[width=4.5cm]{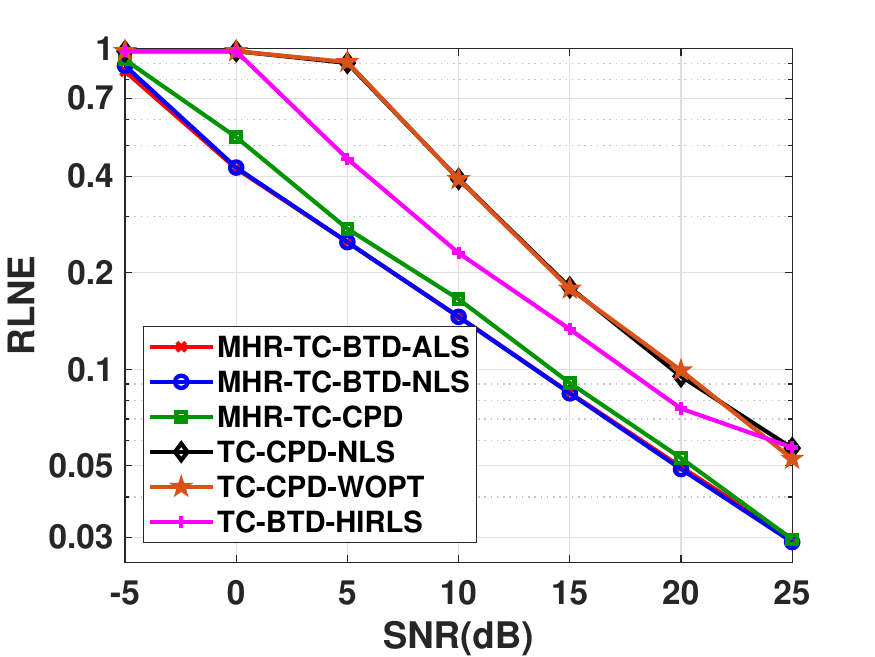}}
  \vspace{-2.6mm}
  \caption{RLNE versus SNR.}
  \label{pic:RLNEversusSNR}
\end{minipage}
\begin{minipage}[b]{0.495\columnwidth}
  \centering
  \centerline{\hspace{3mm}\includegraphics[width=4.5cm]{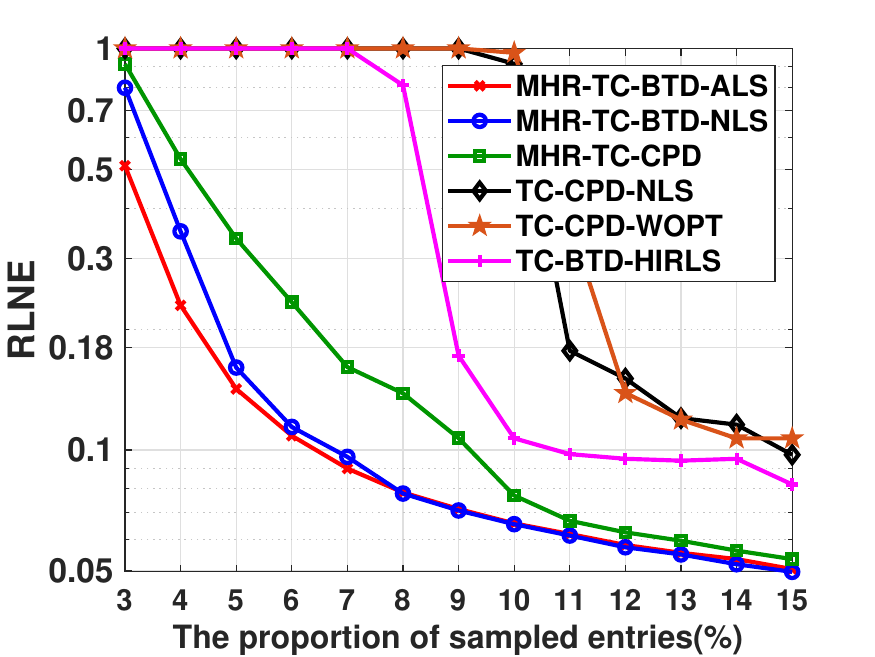}}
  \vspace{-2.6mm}
  \caption{RLNE versus $\rho$.}
  \label{pic:RLNEversusSampleR}
\end{minipage}
\label{fig:res}
\vspace{-7.5mm}
\end{figure}

In the first experiment, we set $I\!=\!J\!=\!K\!=\!20,\!R\!=\!3,L_1\!=\!L_2\!=\!L_3\!=\!3,\rho\!=\!15\%$ ($F\!=\!9$ for CPD based algorithms), and let SNR vary from $-5$dB to $25$dB. The RLNE versus $\text{SNR}$ curves, with each point calculated as the average over $50$ Monte Carlo runs, are plotted in Fig.$\!$ \ref{pic:RLNEversusSNR}.
The results show that the proposed algorithms perform the best, benefitting from the exploitation of both the inherent low-rank properties of harmonics and the BTD model structure of the observed data tensor. The MHR-TC-CPD algorithm, which also exploits the harmonic properties but adopts a less flexible CPD model, is slightly less accurate. The other competitors, without using the harmonic properties, perform significantly worse.

In the second experiment, we set $I\!=\!J\!=\!K\!=\!20, R\!=\!3, L_1\!=\!L_2\!=\!L_3
\!=\!3$ ($F\!=\!9$ for CPD based algorithms), fix SNR to $20$dB, and let $\rho$ vary from $3\%$ to $15\%$. The RLNE versus $\rho$ curves, with each point calculated as the average over $50$ Monte Carlo runs, are plotted in Fig.$\!$ \ref{pic:RLNEversusSampleR}. The results indicate the proposed are more accurate and can recover the full tensor with fewer observations than other algorithms. In addition, the ALS-based algorithm outperforms the NLS-based counterpart when $\rho$ is smaller than $8\%$, as it tends to obtain more accurate solutions than the NLS-based one within a single iteration under these low proportion of sampled
entries conditions.

\subsection{Wireless CSI Completion}
\ignore{
According to 3GPP standards \cite{3GPP}, MIMO wireless channel data can be modeled using a third-order BTD model as described in \eqref{eq:def_vdm_btd} under some conditions, where $R$ represents , reflecting the spatial consistency of delays. The three dimensions of $\Tcalbf$ correspond to the physical antenna array, time, and subcarrier dimensions.}
We employ the widely recognized channel simulation software QuaDRiGa v2.0.0 \cite{QuaDRiGa_v2}, which supports the channel model in standard 3GPP 38.901 communication protocol \cite{3GPP}, to generate the experimental data. The key parameters include: a single user with one antenna moving at $1$km/h, a dual-polarized $4\times 4$ antenna array at the base station, $16$ time samples with $5$ms intervals, and $100$ uniformly spaced subcarriers. The resulting wireless Sub-6GHz CSI $\Tcalbf$ in ``array-time-frequency'' domain is of size $32\times 16\times 100$.\ignore{ and is with $21$ paths in total.}
\ignore{
\begin{minipage}{\columnwidth}
    \centering
    \begin{tabularx}{\textwidth}{|X|X|X|}
        \hline
        Header 1 & Header 2 & Header 3 \\
        \hline
        Data 1 & Data 2 & Data 3 \\
        Data 4 & Data 5 & Data 6 \\
        \hline
    \end{tabularx}
    \captionof{table}{aaa}
\end{minipage}
}

\begin{figure}[htb]
  \vspace{-3.6mm}
  \centering{\includegraphics[width=4.5cm]{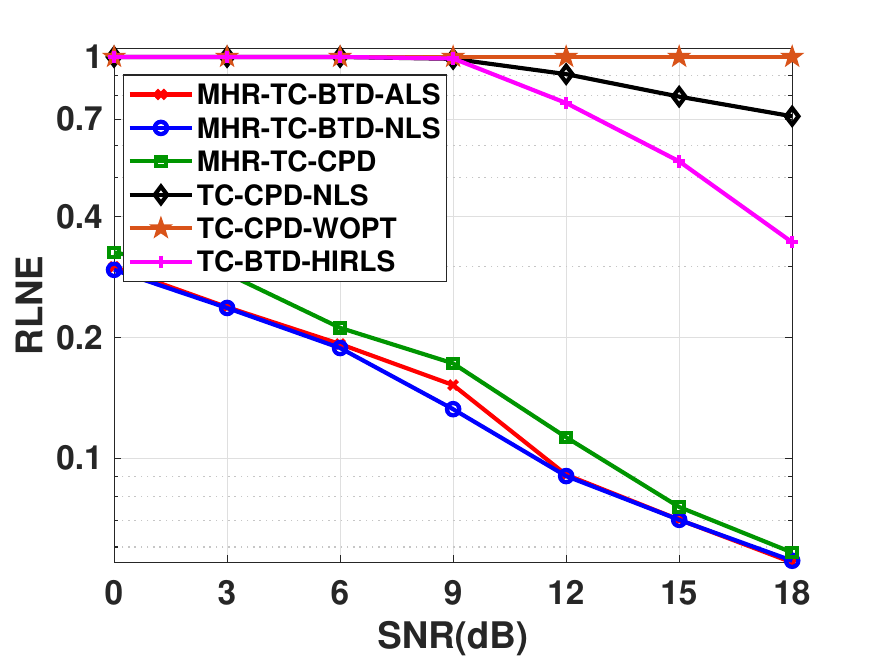}}
  \vspace{-1.9mm}
  \caption{Curves of average RLNE versus SNR on CSI completion.}
  \label{pic:ChannelCompletion}
  \vspace{-4.0mm}
\end{figure}

We observe $\rho=5\%$ entries and set the MLR-$(L_r, L_r ,1)$ BTD model order parameters as $R\!=\!7,L_1\!=\!\cdots\!=\!L_{7}\!=\!3$. The RLNE versus SNR curves, with each point calculated as the average over $50$ Monte Carlo runs, are plotted in Fig.$\!$ \ref{pic:ChannelCompletion}. We have observed in Fig.$\!$ \ref{pic:ChannelCompletion} analogous results as those in Fig.$\!$ \ref{pic:RLNEversusSNR} and Fig.$\!$ \ref{pic:RLNEversusSampleR}, with the proposed algorithms having the best performance, followed by the MHR-TC-CPD algorithm, for similar reasons discussed in the first experiment. Besides, we note that TC-BTD-HIRLS outperforms TC-CPD-NLS and TC-CPD-WOPT as its leverage of the BTD structure.

\ignore{\hl{, especially in high-noise scenarios, where distinguishing harmonic components in the frequency domain becomes increasingly challenging}} \ignore{The MHR-TC-CPD algorithm is slightly less accurate due to the exploitation of the of harmonics structure on the CPD model. The TC-BTD-HIRLS algorithm can be available under small noise benefitting from
the exploitation of the BTD structure. The TC-BTD-HIRLS is better than TC-CPD-NLS and TC-CPD-WOPT for it take use if the BTD structure. other competitors are not effective in this MHR task due to similar reasons.} 

\section{Conclusion}
\label{sec:conclusion}
We propose a tensor completion method of MH signals based on the MLR-$(L_r,L_r,1)$ BTD model, which captures the ``one to 
many" correspondence among harmonics across different modes. We formulate a multi-objective optimization problem that consists of two subproblems of BTD structure fitting and rank minimization of hankelized harmonics, and propose ADMM based algorithms for its computation. Experimental results show that the proposed algorithms offer better performance than existing tensor completion algorithms, with regards to improved accuracy in noisy applications, and fewer observed data rate. Its applicability in a practical application of wireless Sub-6GHz CSI completion is validated.

\vfill
\pagebreak
\bibliographystyle{IEEEbib}
\bibliography{refs}

\end{document}